\documentclass[preprint,preprintnumbers,amsmath,amssymb,aps,pra]{revtex4}

\usepackage{amsmath}
\usepackage{amssymb}
\usepackage{graphics}
\usepackage{graphicx}

\begin{document}
\title{Nonlinear dynamics of a semiquantum Hamiltonian in the vicinity of quantum unstable regimes}
\author{A.M. Kowalski, R. Rossignoli} \affiliation{Departamento de F\'{\i}sica-IFLP,
Universidad Nacional de La Plata-CIC \\
 C.C.67, La Plata (1900), Argentina}
\begin{abstract}
We examine the emergence of chaos in a  non-linear model derived from a semiquantum Hamiltonian describing the coupling between a classical field and a  quantum system. 	
The latter corresponds to a bosonic version of a BCS-like Hamiltonian, and  possesses  stable and unstable regimes.  The dynamics of the whole system is shown to be strongly influenced by the quantum subsystem. In particular, chaos is seen to arise in the vicinity of a quantum critical case, which separates the stable and unstable regimes of the bosonic system.
	
\end{abstract}
 \maketitle

\section{Introduction}
The interplay between quantum and classical systems is a topic of great interest in quantum dynamics and quantum chaos. Whenever quantum effects in one of the systems are negligible in comparison with those of the other, its consideration as classical simplifies the description and provides deep insight into the combined system dynamics.  Examples can be readily found, such as  Bloch equations \cite{Bloch}, two-level systems interacting with an electromagnetic field within a cavity and  Jaynes-Cummings semi-classical model \cite{Milonni,Sa.91}, collective nuclear motion \cite{Ring}, etc. Here we will consider  a semiquantum bipartite system in which the quantum component, representing the matter and described by a quadratic Hamiltonian  in boson operators or generalized coordinates and momenta,  can  exhibit distinct dynamical regimes \cite{RK.05,RK.09} (bounded or unbounded), whereas the classical component represents a single mode of an electromagnetic field. Such type of composite system is of interest in Quantum Optics and Condensed Matter \cite{Milonni,Sa.91,K0,K1}. The essential point we want to discuss is how the different regimes of the quantum system influence those of the combined semiquantum system, and in particular examine if the onset of chaos can be related to this effect.

As is well known, quadratic Hamiltonians in generalized coordinates and momenta, or equivalently in boson operators, are a common presence in theoretical models of physical
systems. They often emerge through diverse linearization procedures  of the pertinent equations of motion around a stationary point \cite{Ring,BR.86}, providing a tractable description of the small amplitude fluctuations  which is exact if the deviations from equilibrium are sufficiently small. They  play in particular a fundamental role in the description of  Bose-Einstein condensates (BEC)
\cite{GM.97,PB.97,AK.02,F.07,E.07,F.08} as well as in other fields like disordered systems \cite{GC.02}, quantum optics \cite{Sa.91},  dynamical systems
\cite{P.97, Do.00, Ko.00} and collective nuclear motion \cite{Ring}.  While the treatment of such quadratic systems in the stable regime is of course standard, leading to a set of normal coordinates which evolve independently,
that of the unstable regime, in which the quadratic Hamiltonian  is no longer positive definite, is  less trivial and requires the introduction of non-hermitian normal coordinates (complex normal modes) \cite{RK.05, RK.09,RRC.14}, which exhibit exponential evolutions. Moreover, at the boundary between stable and unstable sectors,  non-separable regimes  can arise in which the dynamics is described by non-diagonalizable evolution matrices and the equations of motion cannot be fully decoupled \cite{RK.09}.

Non-positive quadratic bosonic forms  naturally emerge  in the description of BEC instabilities \cite{F.07,E.07,F.08}
and fast rotating condensates \cite{D.05,BDS.08,Ft.01,Ft.07,O.04,A.09}, as well
as in generalized RPA treatments \cite{A.97,RC.97}. The  methods developed in \cite{RK.05} were
used  for describing the onset of instabilities in trapped BEC's with a highly quantized vortex
\cite{F.07,E.07,F.08} through the Bogoliubov-de Gennes equations.
Here we will apply this methodology for studying the dynamics of a quantum systems interacting with a classical system,  showing that non-diagonalizable regimes can be  related to the onset of chaos.

\section{The model}

We consider a semi-quantum system composed of two quantum harmonic modes coupled to a classical oscillator which represents a single-mode of an electromagnetic field.
The complete Hamiltonian is of the form
\begin{equation}\label{Hparti}
H=\varepsilon_+(b_{+}^{\dagger }b_{+} + \frac{1}{2})+\varepsilon_-(b_{-}^{\dagger }b_{-}+ \frac{1}{2}) + (\Delta + \alpha X) \,(b_{+} b_{-}+b_{-}^{\dagger}b_{+}^{\dagger}) + \frac{\omega }{2}(P_X^{\,\,\,2}
+X^{\,2}),
\end{equation}
where $b^\dagger_\pm$, $b_{\pm}$ are boson creation and annihilation operators satisfying the standard
commutation relations ($[b_\mu,b^\dagger_{\nu}]=\delta_{\mu\nu}$, $[b_\mu,b_\nu]=[b^\dagger_{\mu},b^\dagger_{\nu}]=0$ for $\mu,\nu=\pm$), $\varepsilon_{\pm}>0$ are the single boson energies  and $X$, $P_X$ are classical coordinate and momentum variables, with $\omega$ the corresponding oscillator frequency.

The dynamical equations for the quantum observables are the canonical ones \cite{K0,K1},
i.e., any operator $O$ evolves  in the Heisenberg picture  as
\begin{equation}
 i \frac{d O }{dt} = - [\: H,  O \:]\,.
\label{Eccanoncero}
\end{equation} The concomitant evolution equation for its  mean value
$\langle O\rangle\equiv {\rm Tr}\,[\rho\,O(t)]$ is
\begin{equation}
i \frac{d \langle  O \rangle}{dt} = - \langle[\:  H,
  O \:]\rangle,
\label{Eccanon}
\end{equation}
where the average is taken with respect to a proper quantum density operator $\rho$.
Additionally,  the classical variables obey classical Hamiltonian equations of motion, i.e.,
\begin{subequations}
\label{eqclasgen}
\begin{eqnarray}
\frac{dX}{dt} & = & \frac{\partial \langle  H \rangle}{\partial P_X}
\label{ds},\\
\frac{dP_X}{dt} & = & - \frac{\partial \langle  H \rangle}{\partial X}. \label{clasgenb}
\end{eqnarray}
\end{subequations}

The complete set of equations (\ref{Eccanon}) + (\ref{eqclasgen})  constitute an autonomous set of coupled first-order ordinary differential equations (ODE). They allow for a dynamical description in which no quantum rules are violated, i.e., the commutation-relations are trivially conserved for all times, since
the quantum  evolution is the canonical one for an effective time-dependent Hamiltonian ($X$ plays the role of a time-dependent parameter for the quantum system) and the initial conditions are determined by a proper quantum density operator $\rho$.

Defining the hermitian operators
\begin{eqnarray}N&=&b_{+}^{\dagger}b_{+}+b_{-}^{\dagger}b_{-}\,,\;\;\;\delta N=b_{+}^{\dagger}b_{+}-b_{-}^{\dagger}b_{-}\,,\\
O_{+}&=&b_{+} b_{-}+b_{-}^{\dagger}b_{+}^{\dagger}\,,\;\;O_{-}=i(b_{+} b_{-}-b_{-}^{\dagger}b_{+}^{\dagger})\,,
\end{eqnarray}
we can rewrite the Hamiltonian (\ref{Hparti})  as
\begin{equation}\label{Hparti2}
H=\varepsilon \, (N+1)+\gamma\, \delta N + (\Delta+\alpha X)\,O_{+} +\frac{\omega }{2}(P_X^{\,\,\,2}
+X^{\,2}),
\end{equation}
where $\varepsilon=(\varepsilon_++\varepsilon_-)/2>0$ and $\gamma=(\varepsilon_+-\varepsilon_-)/2$,
with $|\gamma|<\varepsilon$.
Using Eqs.\ (\ref{Eccanon})--(\ref{eqclasgen}) we then obtain the following closed system of equations for the previous set of
operators and classical variables:
\begin{subequations}
	\label{eqquant1}
	\begin{eqnarray}
	\frac{d\langle N+1\rangle }{dt} &=&2(\Delta + \alpha X) \langle O_{-}\rangle , \label{8a} \\
	\frac{d\langle O_{-}\rangle }{dt} &=&2(\Delta + \alpha X) \, \langle N+1\rangle  +2 \varepsilon\langle
	O_{+}\rangle , \label{8b}\\
	\frac{d\langle O_{+}\rangle }{dt} &=&- 2 \varepsilon \langle
	O_{-}\rangle ,\label{8c}\\
\frac{dX}{dt} &=&\omega P_X,\label{8d} \\
\frac{dP_X}{dt} &=&-(\omega X+\alpha \langle O_{+}\rangle) \label{8e}
	\end{eqnarray}
\end{subequations}
with $d\langle \delta N\rangle/dt=0$.

Eqs.\ (\ref{eqquant1})  constitute  a nonlinear closed ODEs system. Nonlinearity is introduced by the coupling factor between the two systems controlled by the parameter $\alpha$. For $\alpha=0$
the two systems become decoupled and previous equations reduce, accordingly, to two independent {\it linear} systems.

The mean value $\langle O_{-}\rangle $
represents a ``current'' while $\langle O_{+}\rangle $ determines the
expectation value of the quantum part of the interaction
potential. Each level population
can be recovered as $\langle b^\dagger_{\pm}b_{\pm}\rangle=(\langle N\rangle\pm\langle\delta N\rangle)/2$.
The full system (\ref{eqquant1})  possesses in addition   the following Bloch-like  invariant of motion, \begin{equation}\label{Inv}
I = \langle N+1 \rangle^{2} - 4|\langle b_+b_-\rangle|^2= \langle N+1 \rangle^{2}-\langle  O_{-} \rangle^{2} - \langle  O_{+} \rangle^{2},
\end{equation}
which satisfies $dI/dt=0$  in both the linear ($\alpha=0$) and nonlinear ($\alpha\neq 0$) cases, as can be directly verified.

The conservation of $\langle \delta N\rangle$ makes  it convenient to
work with the effective energy $E_{\rm eff}=\langle H \rangle-\gamma \, \langle \delta N \rangle -\varepsilon$
instead of the total energy $\langle H \rangle$. Both quantities are invariants of motion.
Using $I$ together with the effective energy, we can reduce the original number of degrees of freedom
of the system (\ref{eqquant1}) to three. This property allows us to use tools like the  {\it Poincare sections} to analyze the system dynamics.

In the linear case $\alpha=0$, the evolution of the quantal  subsystem is fully determined  by the quantum Hamiltonian
\begin{equation}
H_q=\varepsilon_+(b_{+}^{\dagger }b_{+} + \frac{1}{2})+\varepsilon_-(b_{-}^{\dagger }b_{-}+ \frac{1}{2}) + \Delta\,(b_{+} b_{-}+b_{-}^{\dagger}b_{+}^{\dagger})\,.
\end{equation}
The ensuing dynamics exhibits  {\it three}  distinct regimes according to the value of the coupling strength $\Delta$ \cite{RK.05}: \\
a) The dynamically stable regime,  which holds for $|\Delta|<\varepsilon$,
where the evolution is {\it bounded and quasiperiodic}. Here $H_q$ can be written as a sum of two independent standard normal modes
\begin{equation} H_q=\lambda_+(a^\dagger_+a_++\frac{1}{2})+\lambda_-(a^\dagger_-a_-+\frac{1}{2})\,,
\end{equation}
where the eigenfrequencies $\lambda_{\pm}$ are real and given by
\begin{equation}\lambda_{\pm}=\sqrt{\varepsilon^2-\Delta^2}\pm \gamma \,.\label{lambda}
\end{equation}
Here $a_{\pm}=ub_{\pm}+vb_{\mp}^\dagger$, $a^\dagger_{\pm}=u b^\dagger_{\pm}+v b_{\mp}$,
are normal boson creation and annihilation operators ($[a_{\mu},a^\dagger_{\nu}]=\delta_{\mu\nu}$,
$[a_\mu,a_\nu]=[a^\dagger_{\mu},a^\dagger_{\nu}]=0$) related to the original ones through a Bogoliubov transformation \cite{Ring} (with $u=\sqrt{\frac{\varepsilon+\eta}{2\eta}}$, $v=\sqrt{\frac{\varepsilon-\eta}{2\eta}}$ and $\eta=\sqrt{\varepsilon^2-\Delta^2}$ real). Their
evolution is then given by Eq.\ (\ref{Eccanoncero}), i.e., $i d a_{\pm}/dt=\lambda_{\pm} a_{\pm}$,
which leads to
$a_{\pm}(t)=e^{-i\lambda_{\pm}t}a_{\pm}(0)$ and hence  $a^\dagger_{\pm}(t)=e^{i\lambda_{\pm}t}a^\dagger_{\pm}(0)$.

This regime can actually be divided in three subregimes according to the spectrum of $H$ \cite{RK.05}, which in this case is  {\it discrete}, i.e.,    $E_{mn}=\lambda_+(m+\frac{1}{2})+\lambda_-(n+\frac{1}{2})$, with
$m,n\in\mathbb{N}$:\\
 a$_1$) $|\Delta|< \sqrt{\varepsilon_+\varepsilon_-}=\sqrt{\varepsilon^2-\gamma^2}$,
where $\lambda_{\pm}$ are both {\it positive} and $H$ is then {\it positive definite}; \\
a$_2$) $|\Delta|=\sqrt{\varepsilon^2-\gamma^2}$, where $\lambda_+>0$ but $\lambda_-=0$,
implying $H$ positive {\it semidefinite} with a discrete yet infinitely degenerate spectrum; \\
a$_3$) $\sqrt{\varepsilon^2-\gamma^2}<\Delta<\varepsilon$, where $\lambda_+>0$ but $\lambda_-<0$,
entailing that $H$ is non longer positive and has no longer a  minimum energy.
\\

b) The dynamically unstable  regime,  existing for  $|\Delta|>\varepsilon$, where the dynamics is {\it exponentially unbounded}. Here $H_q$ can be written as a sum of two complex normal modes \cite{RK.05},
\begin{equation} H_q=\lambda_+(\bar{a}_+a_++\frac{1}{2})+\lambda_-(\bar{a}_-a_-+\frac{1}{2})\,,
\end{equation}
where $\lambda_{\pm}$ are still given by Eq.\ (\ref{lambda}) but are now {\it complex}, and where
 $a_{\pm}=u b_{\pm}+v b_{\mp}^\dagger$, $\bar{a}_{\pm}=u b^\dagger_{\pm}+ v b_{\mp}$,
 with $u,v$ given by the same previous expressions,
 still satisfy boson commutation relations ($[a_\mu,\bar{a}_{\nu}]=\delta_{\mu\nu}$,
 $[a_\mu,a_\nu]=[\bar{a}_\mu,\bar{a}_\nu]=0$) but $\bar{a}_{\pm}\neq a^\dagger_{\pm}$, since
 $u,v$ are now also complex (complex normal modes). These operators then exhibit,
 according to Eq.\ (\ref{Eccanoncero}), exponential-type evolutions $a_{\pm}(t)=e^{-i\lambda_{\pm}t}a_{\pm}(0)$, $\bar{a}_{\pm}(t)=e^{i\lambda_{\pm}t}\bar{a}_{\pm}(0)$,
 which diverge either for $t\rightarrow \infty$ or $t\rightarrow-\infty$.
 Note that hermiticity is preserved, since $\lambda_\pm^*=-\lambda_{\mp}$ and
 $(\bar{a}_\pm a_{\pm})^\dagger=-a_{\mp}\bar{a}_{\mp}$.

c) The non-separable case  $|\Delta|=\varepsilon$, where $\lambda_\pm=\pm\gamma$ and $H$ can no longer be written as a sum of two-independent modes. This case, which lies at the border between the dynamically stable and unstable regimes, corresponds to a non-diagonalizable evolution matrix \cite{RK.05} and hence to a linear system {\it which cannot be fully decoupled}. Instead, $H_q$ can be written here as
\begin{equation}
H_q=\gamma\,(\bar{a}_+a_+-\bar{a}_-a_-)+2\Delta\bar{a}_-\bar{a}_+\,,
\end{equation}
where $a_{\pm}=\frac{b_{\pm}-b^\dagger_{\mp}}{\sqrt{2}}$, $\bar{a}_{\pm}=\frac{b_{\pm}-b^\dagger_{\mp}}{\sqrt{2}}$ still satisfy boson-like commutation relationships. In this form, $H_q$ is ``maximally decoupled'', in the sense that the evolution equations for $\bar{a}_{\pm}$ are fully decoupled, while those for $a_{\pm}$ are coupled just to $\bar{a}_{\mp}$. This leads to $\bar{a}_{\pm}(t)=e^{\pm i\gamma t}\bar{a}_{\pm}(0)$,
	$a_{\pm}(t)=e^{\mp i\gamma t}[a_{\pm}(0)-2it\Delta \bar{a}_{\mp}(0)]$, and hence to a polynomially unbounded evolution \cite{RK.05}.

Previous expressions for $a_{\pm}(t)$ and $\bar{a}_{\pm}(t)$ allow one to obtain the final explicit expressions for the averages of the  relevant observables. In the diagonalizable cases a)  and b), we obtain 
\begin{eqnarray}
\langle N+1\rangle&=&-\frac{-\varepsilon(\varepsilon \langle N+1\rangle_0+\Delta\langle O_+\rangle_0) + 	\Delta (\Delta\langle N+1\rangle_0 + \varepsilon\langle O_+\rangle_0) \cos 2\eta t - \Delta\eta\langle O_-\rangle_0
	\sin 2\eta t}{\eta^2},\nonumber\\&&\\
\langle O_-\rangle&=&\langle O_-\rangle_0 \cos 2\eta t +
(\Delta\langle N+1\rangle_0+\varepsilon \langle O_+\rangle_0)\frac{\sin 2\eta t}{\eta},\\
\langle O_+\rangle&=&-\frac{\Delta (\varepsilon\langle N+1\rangle_0+\Delta \langle O_+\rangle_0) -
	\varepsilon (\Delta\langle N+1\rangle_0+\varepsilon\langle O_+\rangle_0)
	\cos 2 \eta t + \varepsilon\eta\langle O_-\rangle_0\sin 2\eta t}{\eta^2},	\nonumber\\&&
\end{eqnarray}
where $\eta=\sqrt{\varepsilon^2-\Delta^2}$  is real for $|\Delta|<\varepsilon$ (case a) and imaginary for $|\Delta|>\varepsilon$ (case b).
In the non-diagonalizable transition regime  $|\Delta|=\varepsilon$ ($\eta=0$), the explicit expressions become
\begin{eqnarray}
\langle N+1\rangle&=&\langle N+1\rangle_0 + 2\langle O_-\rangle_0 \varepsilon t + 2
\langle N+1+O_+\rangle_0 \varepsilon^2 t^2,\label{N+1 nond} \\
\langle O_-\rangle&=&\langle O_-\rangle_0 + 2\langle N+1+O_+\rangle_0 \varepsilon t,\\
\langle O_+\rangle&=&\langle O_+\rangle_0 - 2\langle O_-\rangle_0 \varepsilon t -
2\langle N+1+O_+\rangle_0 \varepsilon^2 t^2.
\end{eqnarray}

\section{Results}
\begin{figure}[ht!]
	\centerline{
		\scalebox{1.}{\includegraphics{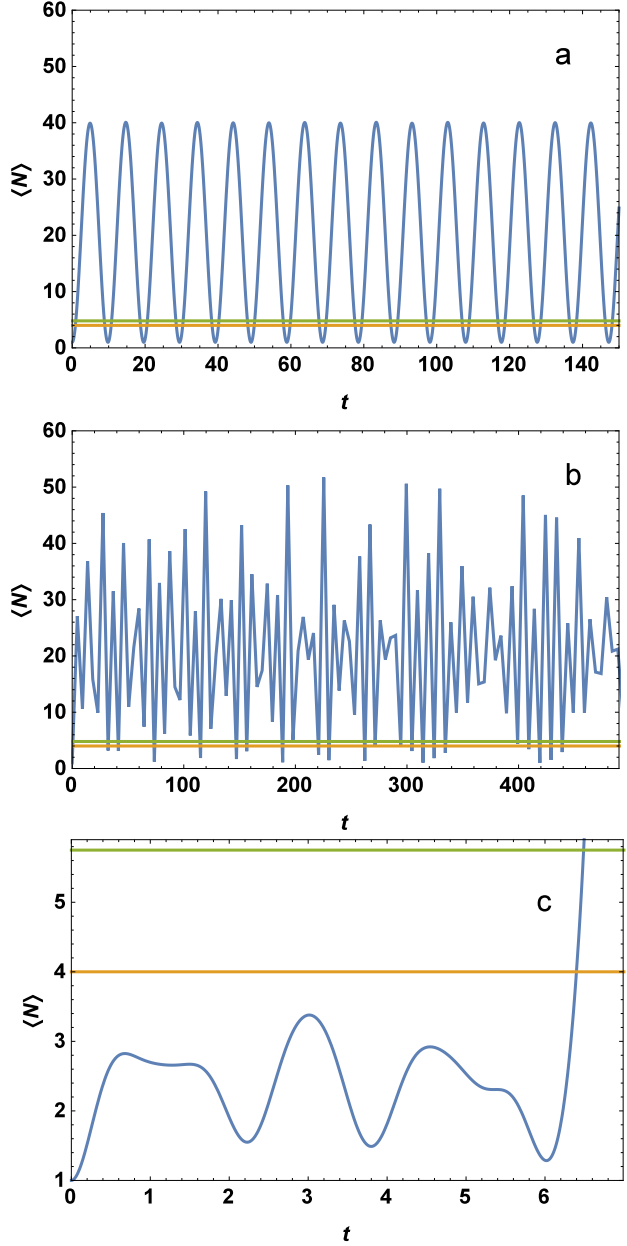}}}
	\caption{Plot of $\langle N \rangle$ (blue solid lines) and the invariants $E_{\rm eff}$ (green lines)  and $I$ (orange lines) for the initial conditions
		$\langle N \rangle_0=1$, $\langle O_{\pm} \rangle_0=0$,  $X_0=1$ and $P_0=-2.54950976$, with parameters $\omega=1$ and $\alpha/\Delta=0.0001$ (a), 0.015 (b) and 1.1 (c), while  $\varepsilon/\Delta=1.05$ (a)--(b), and $2$ (c). Fig.\ 1a corresponds to the oscillatory zone and 1b to the non-linear and chaotic one. In 1c,  $\langle N \rangle$  diverges for large $t$.} \label{Fig1}
\end{figure}

\begin{figure}
	\centerline{\hspace*{.5cm}\scalebox{.75}{\includegraphics{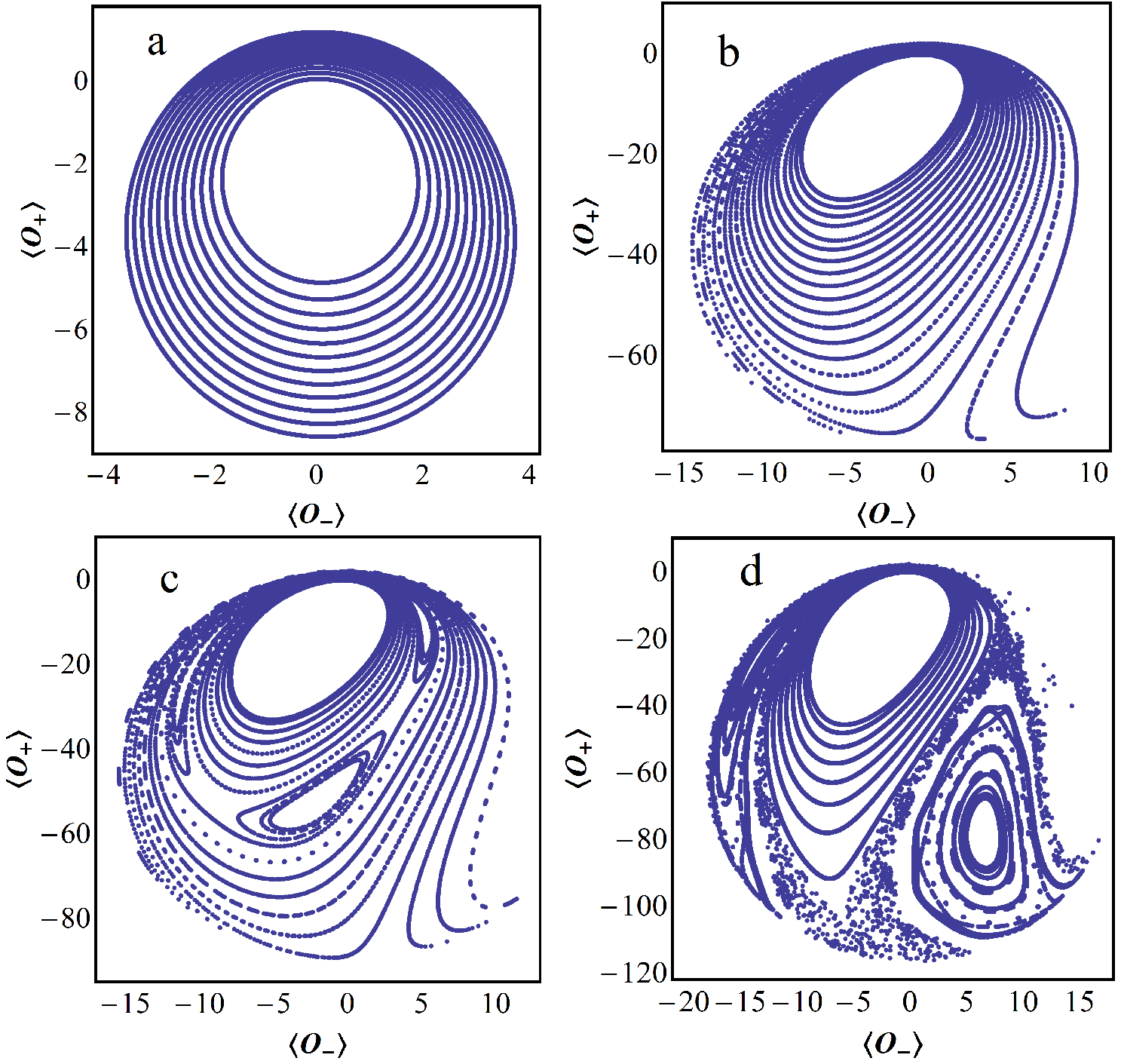}}}
	\caption{Poincare sections $\langle O_{+} \rangle$ vs. $\langle O_{-} \rangle$ for $X=0$,
		corresponding to  $E_{\rm eff}=4.8$ and $I=4$, with $X_0=1$, $\langle N \rangle_0=1$, $\langle O_{+} \rangle_0=0$ and $\omega=1$. The values of $\langle O_{-} \rangle_0$ and $P_0$ change in  order to generate the 21 curves. We set  $\alpha/\Delta=0.015$ while  the ratios  $\varepsilon/\Delta$ are given by  (a) $\varepsilon/\Delta=1.5$, (b) $\varepsilon/\Delta=1.075$, (c) $\varepsilon/\Delta=1.065$, (d) $\varepsilon/\Delta=1.05$. For decreasing values of  $\varepsilon/\Delta$, the behavior evolves from periodic curves to complex quasiperiodic curves and finally to chaos.}\label{Figs2}
\end{figure}

The numerical results were obtained for  initial conditions consistent with a
proper density operator, such that the pertinent uncertainty relations are satisfied for all times. We have also checked their accuracy by verifying the constancy in time of the dynamical invariants $E_{\rm eff}$ and $I$ (within a precision of $10^{-10}$). 

The obtained numerical results indicate that the distinct regimes obeyed by the semiclassical system are determined by the relation between $\varepsilon$, $\Delta$ and $\alpha$ irrespective of the initial conditions and the value of  $\omega$. When $|\alpha|\geq\varepsilon$, the dynamics is always divergent (with or without oscillations), as occurs in the linear case, with $\alpha$ playing the role of $\Delta$. In Fig.\ 1 we show a characteristic evolution of $\langle N \rangle$ (together with those of the invariants, as check).  When $\varepsilon > |\alpha|$,
the dynamics is determined by  $\varepsilon$, $\Delta$ y $\alpha$, competing $\varepsilon$ with the two coupling constants, but as $\alpha$ decreases,
the system approaches the linear case and hence the relation between $\Delta$ and $\varepsilon$
becomes dominant. In Figs. 2 and 3 we show the Poincare sections obtained from the $X(t)=0$ plane,
for the same values of $E_{\rm eff}$ and $I$.  In Figs.\ 2,   $\alpha< \varepsilon$ is kept fixed but the ratio $\varepsilon/\Delta$ is varied. It is seen that for $\varepsilon > |\Delta|$ the dynamics is periodic  (Fig.\ 1a and Figs.\ 2a-2b) as in the linear case for most ratios, but becomes quasiperiodic (Fig.\ 2c) in the vicinity of the non-diagonalizable regime  $\varepsilon=|\Delta|$, showing increasing nonlinear effects as this region is approached (Fig.\ 1b).
For $\varepsilon < |\Delta|$ the regime becomes divergent as in the linear case.

\begin{figure}
	{\scalebox{.75}{\includegraphics{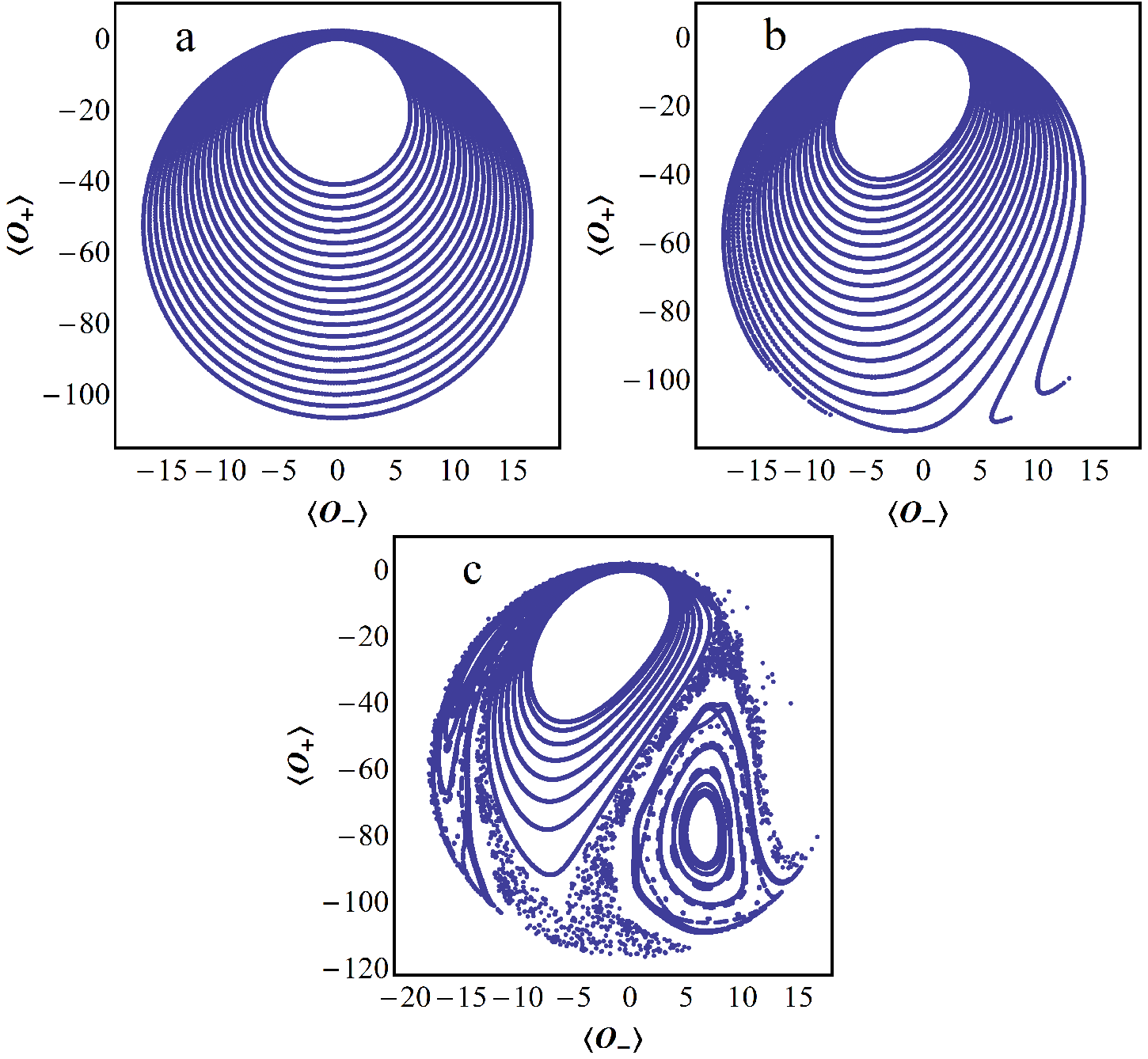}}}
	\caption{Poincare sections $\langle O_{+} \rangle$ vs. $\langle O_{-} \rangle$ for $X=0$ as in Fig.\ 2, with the same inital conditions and parameter values, but setting $\varepsilon/\Delta=1.05$ (fixed) and  varying $\alpha/\Delta$: (a) $\alpha/\Delta=0.0001$, (b) $\alpha/\Delta=0.01$, (c) $\alpha/\Delta=0.015$. For increasing values of $\alpha/\Delta$, the behavior  evolves again from
		periodic curves to complex quasiperiodic curves and finally to chaos. Fig. (c) coincides with
		Fig. 2 (d). }  \label{Figs3}
\end{figure}

\begin{figure}
	\centerline{
		\scalebox{.85}{\includegraphics{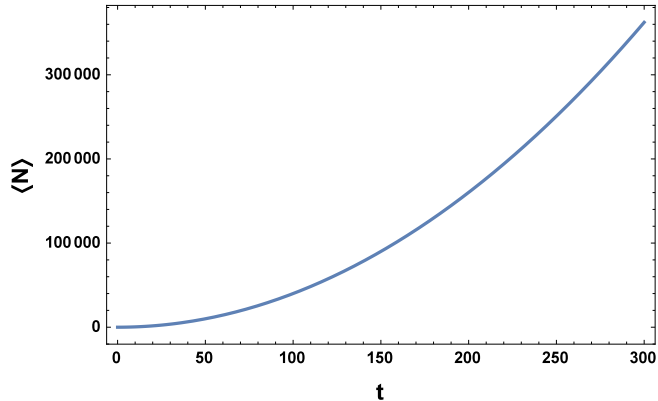}}}
	\caption{$\langle N \rangle$ vs.\ $t$. The initial conditions and parameter values are those of Fig.\ \ref{Fig1}, but now $\varepsilon=\Delta=1$ and $\alpha=10^{-6}$. This is the non-diagonalizable region of the linear case. For the depicted times this curve is practically similar to that of the linear case.} \label{Fig4}
\end{figure}

\begin{figure}
	\centerline{
		\scalebox{1.}{\includegraphics{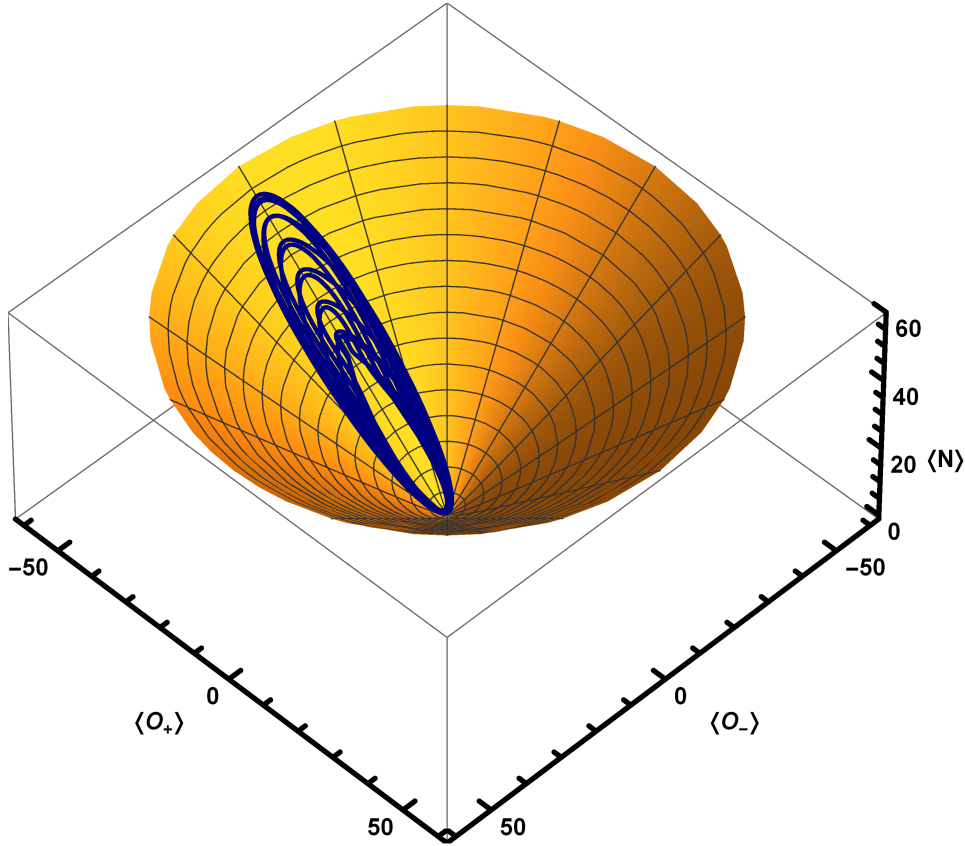}}}
	\caption{Solution of the subsystem (\ref{eqquant1}) for one of the cuasiperiodic curves of Fig. 2d).
		In this case Eq.\ (\ref{Inv}) represents a two-sheet hyperboloid ($I=4$). } \label{Figs5}
\end{figure}

\begin{figure}
	\centerline{\hspace*{.5cm}\scalebox{1.}{\includegraphics{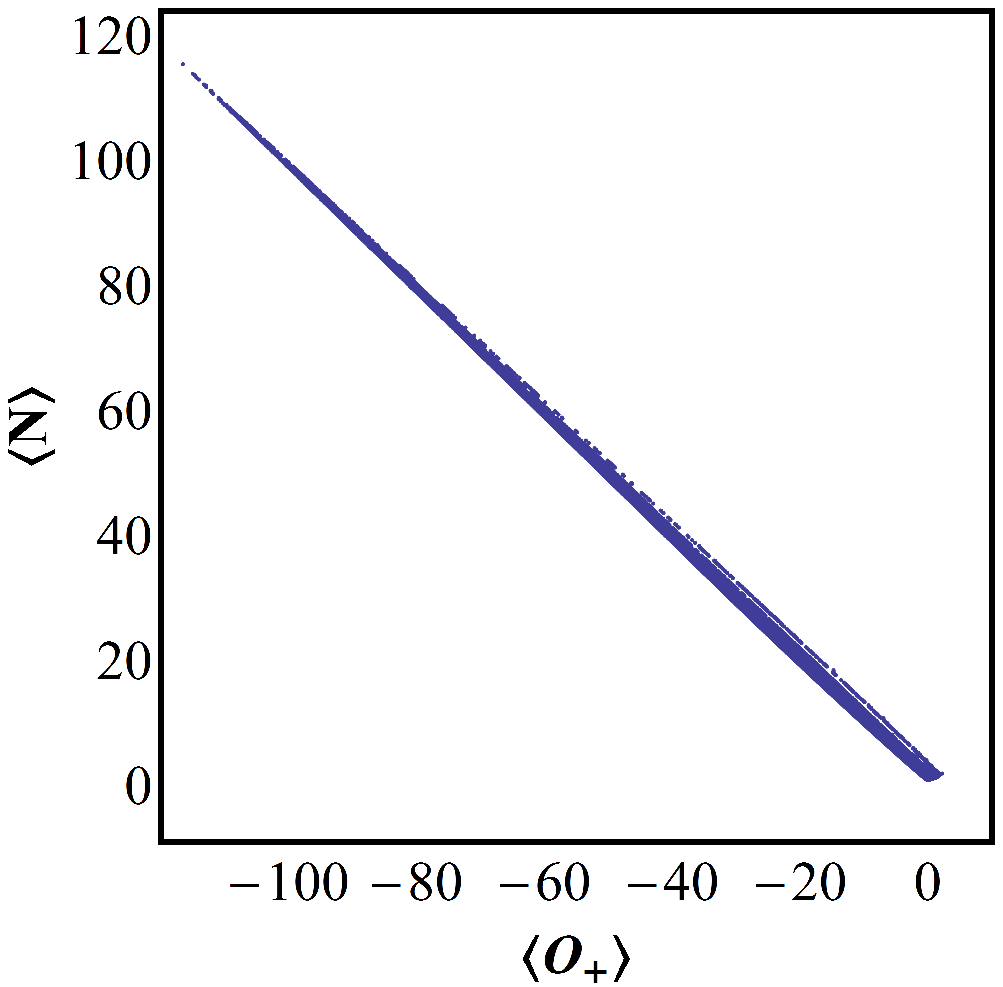}}}
	\caption{Poincare section $\langle N \rangle$ vs.\ $\langle O_{+} \rangle$ for $X=0$.
		This plot corresponds to  Fig.\ 2 (d) (same initial conditions and parameter values). In the nonlinear case it is no longer a surface.} \label{Fig6}
\end{figure}

The most remarkable behavior occurs in the critical regime  $\varepsilon \simeq|\Delta|$,
i.e. in the vicinity of the non-diagonalizable case of the linear system, at border with the unstable case. Here we find complex quasiperiodic evolution curves (Fig. 2c). Moreover, for
appropriate ``small'' values of $\alpha$ ($\alpha<\Delta$), chaos is seen to emerge, as shown in Fig.\ 2d, where
the characteristic presence of chaotic see is recognized.  In Figs. 3, we see the same behavior
for different values of  $\alpha<\varepsilon$, maintaining the same ratio $\varepsilon/\Delta$.
Fig.\ 3c, coincides with Fig.\ 2d. The existence of chaos was verified by the calculation of the  Lyapunov characteristic exponent, which is positive for the curves in question.
On the other hand, the linear regime is approached when $\alpha$
decreases (Fig.\ 4). Here we have  $\varepsilon = \Delta$ for  $\alpha=10^{-6}$. The result
is practically the same as in the linear case (Eq. (\ref{N+1 nond})).

Eq. (\ref{Inv}) represents different types of surfaces in the
space $(\langle N \rangle, \langle O_{-} \rangle, \langle O_{+} \rangle)$. If  $I>0$ it represents a two-sheet hyperboloid whereas
for $I=0$ it is a cone. 
Surfaces are obviously limited by the condition  $\langle N \rangle \geq 0$.

In  Fig.\ 5, we depict a solution of subsystem (\ref{eqquant1}), corresponding to a quasiperiodic curve of Fig. 2d. This solution rests on the intersection of surface $I=4$ and the region $E_{\rm eff}=4.8$. This last quantity represents a plane in the linear case ($\alpha=0$), but in the nonlinear case is no longer a surface, as seen in Fig.\  6. The curves, are no longer plane, and this fact enables the onset of chaos.

\section{Conclusions}

In this article we have studied the dynamics of a semiquantum system resulting from the interaction of a bosonic system with a classical field. 
The dynamics of the bosonic version of a BCS-like pairing Hamiltonian was solved with a methodology \cite{RK.05,RK.09}  suitable for completely general, not necessarily positive,  quadratic forms. Through this methodology we found the existence of a quasi-periodic regime, a divergent regime and an intermediate regime that corresponds to a non-diagonalizable (i.e., non-separable) case. These zones are determined by the relation between the parameter $\varepsilon$ and the quantum coupling constant $\Delta$ ( $\varepsilon > |\Delta|$, $\varepsilon < |\Delta|$, $\varepsilon = |\Delta|$ respectively) \cite{RK.05}.

This quantum Hamiltonian is coupled  to a classical harmonic oscillator that represents a mode of the electromagnetic field. The coupling introduces nonlinear effects in the equations of motion. The nonlinear dynamics of the composite system is determined by $\varepsilon$ and the two coupling constants $\Delta$ and $\alpha$. If $\varepsilon \leq|\alpha|$ the dynamic is divergent (Fig. 1). If $\varepsilon > |\alpha|$, it will depend on the relation between the three constants, competing $\varepsilon$ with the two coupling constants. The dynamics  can be periodic,  quasi-periodic and also complex and even chaotic as is observed in Figs. 2-3. In these figures Poincare sections are shown for fixed values of the effective energy $E_{\rm eff}$ and of the invariant of motion $I$, together with the plane $X=0$. 
When $\alpha$ tends to zero, the relationship between $\varepsilon$ and $\Delta$ of the linear case is recovered. In  Fig.\ 4 this situation is observed for a case corresponding to a non-diagonalizable linear regime.

The most remarkable behavior occurs for specific small finite values of $\alpha$.  In this case, in the vicinity of the non-diagonalizable linear regime ($\varepsilon \simeq \Delta$), we can observe the emergence of chaos (Figs.\ 2d -3c).  This result  was tested by the calculation of the  pertinent Lyapunov characteristic exponent, which is positive for these curves.

We can conclude that the use of the aforementioned methodology has facilitated the analysis of the dynamics of the semiquantal nonlinear system through that of the associated linear subsystem. It has also allowed us  to understand the appearance of the chaotic phenomenon by its relation to the non-diagonalizable linear case. Although the presence of the classical system enables the existence of chaos, the previous fact allows us to visualize this effect as a phenomenon emerging from the quantum system.

{\it Acknowledgment}. The authors acknowledge support from CIC and UNLP of Argentina.


\begin{thebibliography}{99}
\bibitem{Bloch} E. Bloch, Phys.\ Rev.\ {\bf 70}, 460 (1946); {\bf 70}, 474 (1946).
\bibitem{Milonni} P. Milonni, M. Shih, J.R.\ Ackerhalt,  {\it Chaos in Laser-Matter Interactions}, World Scientific Publishing Co., Singapore, 1987.
\bibitem{Sa.91} P.~Meystre, M.~Sargent,  {\it Elements of Quantum Optics}, Springer, NY, 1991.
\bibitem{Ring} P. Ring, P. Schuck,  {\it The Nuclear Many-Body Problem}, Springer-Verlag: Berlin, Germany, 1980.
\bibitem{RK.05} R. Rossignoli,  A.M. Kowalski, Phys. Rev. A {\bf 72}, 032101 (2005).
\bibitem{RK.09} R. Rossignoli,  A.M. Kowalski, Phys. Rev. A {\bf 79}, 062103 (2009).
\bibitem{K0} A.M. Kowalski, A. Plastino, A.N. Proto, Phys. Rev. E {\bf 52}, 165 (1995).
\bibitem{K1} A.M. Kowalski, Physica A {\bf 458}, 106 (2016).
\bibitem{BR.86} J.P. Blaizot and G. Ripka, {\it Quantum Theory of Finite
Systems}, MIT Press, MA, 1986.
\bibitem{GM.97} E.V.\ Goldstein, P.\ Meystre, Phys.\ Rev.\ A {\bf 55}, 2935
(1997).
\bibitem{PB.97} H.~Pu, N.P.~Bigelow, Phys.\ Rev.\ Lett.\ {\bf 80}, 1134
(1998);
C.K.~Law, H.\ Pu, N.P.~Bigelow, J.H.~Eberly, Phys.\ Rev.\ Lett.\ {\bf 79},
3105 (1997).
\bibitem{AK.02} S.~Alexandrov, V.V.~Kavanov, J.Phys.\ Condens.\ Matter 14,
L327 (2002); V.I.~Yukalov and E.P.~Yukalova, Laser Phys.\ Lett. 1, 50 (2004);
\bibitem{F.07}E.\ Fukuyama, M.\ Mine, M.\ Okumura, T.\ Sunaga, Y.\ Yamanaka,
Phys.\ Rev.\ A {\bf 76}, 043608 (2007);
M.\ Mine et al, Ann. Phys. {\bf 322}, 2327 (2007).
\bibitem{E.07}T.\ Sunaga et al, J.\ Low Temp Phys.\ {\bf 148}, 381 (2007);
M.\ Mine et al, J.\ Low Temp.\ Phys.\ 148, 331 (2007).
\bibitem{F.08}
Y.\ Nakamura, M.\ Mine, M.\ Okumura, Y.\ Yamanaka,
Phys.\ Rev.\ A {\bf 77}, 043601 (2008).
\bibitem{GC.02} V.~Gurarie, J.T.~Chalker, Phys.\ Rev.\ Lett.\ {\bf 89}, 136801 (2002).
\bibitem{P.97} I.A.\ Pedrosa, Phys.\ Rev.\ A {\bf 55}, 3219 (1997).
\bibitem{Do.00} V.V. Dodonov, J. Phys. A {\bf 33}, 7721 (2000).
\bibitem{Ko.00} A.M.\ Kowalski et al, Phys. Lett. A {\bf 297}, 162 (2002).
\bibitem{RRC.14} L.\ Reb\'on, R.\ Rossignoli, N.\ Canosa, Phys.\ Rev.\ A {\bf 89},
 042312 (2014).
\bibitem{D.05} A.\ Aftalion, X.\ Blanc, J.\ Dalibard,	
Phys.\ Rev.\ {A} {\bf 71}, 023611 (2005); S. Stock et al, Laser Phys.\ Lett. {\bf 2},
275 (2005).
\bibitem{BDS.08}
I.\ Bloch, J.\ Dalibard,   W.\ Zwerger, Rev.\ Mod.\ Phys.\ {\bf 80}, 885 (2008).
\bibitem{Ft.01}
M.\ Linn, M.\ Niemeyer, A.\ L.\ Fetter, Phys.\ Rev.\ A {\bf 64}, 023602 (2001).
\bibitem{Ft.07} A.\ L.\ Fetter, Phys.\ Rev.\ A {\bf 75}, 013620 (2007).
\bibitem{O.04} M.\ \"O.\ Oktel, Phys.\ Rev.\ A {\bf 69}, 023618 (2004).
\bibitem{A.09}A.\ Aftalion, X.\ Blanc, N.\ Lerner,
Phys.\ Rev.\ A {\bf 79} 011603(R) (2009).
\bibitem{A.97} H. Attias, Y. Alhassid, Nucl.\ Phys.\ A {\bf 625}, 363 (1997).
\bibitem{RC.97} R.\ Rossignoli,  N.\ Canosa, Phys.\ Lett.\ B {\bf 394}, 242
(1997);
 R. Rossignoli, N.\ Canosa, P.\ Ring,
 Phys.\ Rev.\ Lett.\ {\bf 80} 1853 (1998); Phys.\ Rev.\ B {\bf 67},
 144517 (2003).


\end{thebibliography}
\end{document}